\documentclass[aps,prb,twocolumn,showpacs,tighten,letterpaper,float,nosuperscriptaddress,footinbib,reprint]{revtex4-1}

\usepackage[pdftex]{graphicx}
\usepackage{dcolumn}
\usepackage{bm}
\usepackage{epsfig}
\usepackage{latexsym}
\usepackage{amsmath}
\usepackage{amsfonts}
\usepackage{amssymb}
\usepackage{float}
\usepackage{caption}
\usepackage{graphicx}
\graphicspath{{images/}}
\usepackage{subcaption}
\usepackage{color}
\usepackage{array}
\usepackage{framed}
\usepackage{soul}

\begin{document}

\title{Fractons from Confinement in One Dimension}

\author{Shriya Pai}
\affiliation{Department of Physics and Center for Theory of Quantum Matter, University of Colorado, Boulder, CO 80309}

\author{Michael Pretko}
\affiliation{Department of Physics and Center for Theory of Quantum Matter, University of Colorado, Boulder, CO 80309}

\date{\today}

\begin{abstract}
Recent work has shown that two seemingly different physical mechanisms, namely fracton behavior and confinement, can give rise to non-ergodicity in one-dimensional quantum many-body systems.  In this work, we demonstrate an intrinsic link between these two mechanisms by studying the dynamics of one-dimensional confining theories, such as a $U(1)$ gauge theory and a quantum Ising model.  We show that, within certain parameter regimes, these models exhibit effective fracton dynamics, characterized by immobility of stable single-particle excitations and free motion of dipolar bound states.  By perturbatively integrating out the linearly confining field, we obtain an effective fracton Hamiltonian for the confined charges which exhibits conservation of dipole moment.  We discuss an intuitive understanding of these results in terms of the motion of the confining strings, leading to potential extensions to higher dimensions.  We thereby interpret recent observations of nonthermal eigenstates and glassy dynamics in confining theories in terms of corresponding results in the fracton literature.
\end{abstract}
\maketitle

\normalsize

\emph{Introduction}.    A major theme in modern condensed matter physics is the study of isolated quantum many-body systems which fail to reach thermal equilibrium, violating the assumptions of classical statistical mechanics.  This is in contrast to the majority of quantum systems which eventually reach an effective thermal state, even when initialized in an eigenstate.  This intuition is codified in the eigenstate thermalization hypothesis (ETH), which holds that all eigenstates of a generic thermalizing quantum system should be locally indistinguishable from thermal states.\cite{Deutsch,Rigol,Srednicki}  The most familiar mechanism leading to breakdown of the ETH is disorder, $i.e.$ a spatially random potential.  For example, it has been known since the seminal work of Anderson that a sufficiently strong disordered potential can localize a system of non-interacting electrons.\cite{anderson}  More recent work has shown that disorder can lead to non-ergodic behavior even in interacting many-body systems.\citep{BAA,GMP,mblarcmp}  This ``many-body localization" (MBL) is typically characterized by a completely nonthermal spectrum, with all states localized.

While studies of non-ergodicity in quantum many-body systems have historically focused on disorder, recent work has shown that nonthermal behavior can occur even in translationally invariant systems, through a variety of mechanisms.  These new mechanisms can lead to unusual incomplete forms of localization, such as quantum many-body scars\cite{Serbyn2,turner2018weak,LinMotrunich,
moudgalya2018entanglement,ShiraishiMori,ChoiAbanin,
ok2019topological,Burnell}, first seen in analytic models such as the AKLT chain \cite{moudgalya2,ShiraishiMori}, which are a small set of localized states coexisting with an otherwise thermalizing spectrum.  One mechanism which can lead to such a partially localized spectrum is fracton behavior.  A fracton is a particle which obeys a set of conservation laws, such as conservation of charge and dipole moment, which prohibits single-particle mobility, but nevertheless allows for motion of certain bound states.\cite{chamon,haah,fracton1,fracton2,sub,fractonarcmp}  (This differs from a previous use of the word ``fracton" in the high energy literature.\cite{khlopov})  This mobility restriction causes fracton systems to generically exhibit glassy dynamics when initialized in a state featuring isolated fractons, such that thermalization occurs only at unreasonably long times.\cite{prem}  In one spatial dimension, it has been shown that certain fracton systems can actually exhibit perfect localization, retaining a memory of initial conditions at arbitrarily long times. \cite{pai2018localization}  This localization manifests in the spectrum of a fracton system as a set of ETH-violating states, such as many-body scars\cite{fractonscars} or fragmentation behavior\cite{ah1,ah2}, depending on certain details.

Along a seemingly separate line of investigation, it has been shown that similar phenomenology can arise in lattice models exhibiting confinement, $i.e.$ with particles interacting via a linearly confining potential, analogous to the behavior of quarks in quantum chromodynamics.  Numerous works have now indicated that such models can exhibit non-ergodic behavior, such as glassy dynamics and many-body scars, without the need for spatial disorder.\cite{kormos,brenes,mazza,glassyquark,crusoe,crusoe2,
holoconf,cubero,lgt,muss1,muss2,muss3,taglia,surace2,
surace3} It has even been argued that the slow dynamics of confining theories is closely related to the long-time oscillations observed in Rydberg atom experiments.\cite{lgt,rydberg}  Upon the introduction of disorder, further interesting phases, such as the Mott glass, are also possible.\cite{yzchou}  Furthermore, in contrast to the naive expectation of a confining theory, these models can exhibit stable configurations of isolated charges which do not collapse back into the vacuum.  At first glance, the phenomenological similarity between fracton systems and confining models may seem purely coincidental.  In this work, however, we establish a concrete link between these two mechanisms for non-ergodic behavior by demonstrating the presence of fracton behavior in certain regimes of typical confining models.

Specifically, we study two well-known one-dimensional lattice models exhibiting confinement: a $U(1)$ lattice gauge theory and a quantum Ising model with transverse and longitudinal fields.  This Ising model can also be equivalently written in terms of a $Z_2$ lattice gauge theory.  In both cases, ``charges" occur at the endpoints of strings, resulting in a linear potential between a pair of charges at opposite ends of a string, $V(R)\sim R$.  (In the Ising case, these charges correspond to isolated domain walls between regions of opposite spin polarization.)  Despite this confining potential, such models have been observed to exhibit stable single-particle excitations.  If the system were in contact with an energy bath, a long string could relax by shrinking in length via motion of its endpoints, or even by generating new charges out of the vacuum.  In isolation, however, such processes violate conservation of energy, with the end result that single-particle motion is suppressed.  Nevertheless, a bound state of two opposite charges, corresponding to a string of fixed length (or a domain of fixed size), can move around free from such energetic restrictions.  This type of behavior, featuring immobile single particles with mobile bound states, is precisely the sort of behavior expected of a fracton system.

In this work, we make this intuition precise by explicitly mapping these lattice gauge theories onto fracton Hamiltonians, at least within certain parameter regimes.  We do so by perturbatively integrating out the confining strings, which yields a dipole-conserving effective Hamiltonian for the charges (see Ref. \onlinecite{sous2019fractons} for an analysis involving similar perturbative techniques to obtain fracton physics in the context of polarons and hole-doped antiferromagnets).  We also discuss how the effective mass of the mobile dipoles increases as a function of the separation of their constituent charges, consistent with the constraints of a local Hamiltonian.  We provide intuition for these results in terms of the motion of the confining strings themselves, which in turn hints at higher-dimensional generalizations.  In light of our results, the presence of non-ergodic behavior in confining models can be equivalently understood in terms of the corresponding results in fracton systems.  Our work will allow for the future productive exchange of ideas between fracton physics and the study of confinement, and may allow for a deeper understanding of the slow dynamics observed in Rydberg atom arrays.\cite{rydberg}

\begin{figure}[t]
 \centering
 \includegraphics[width=0.4\textwidth]{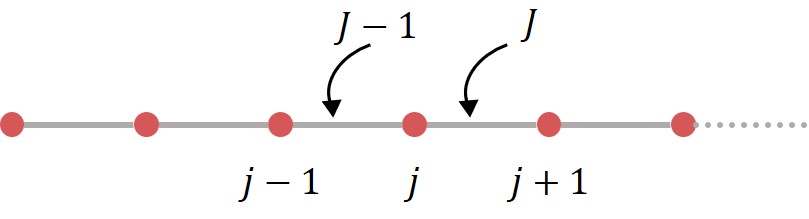}
\captionsetup{justification=raggedright,singlelinecheck=false}
 \caption{\textbf{$U(1)$ lattice gauge theory}. We use $\{j,\ldots \}$ to represent sites and $\{ J, \ldots \}$ to represent links, which host the charges and gauge field respectively.}
 \label{fig:schwinger}
 \end{figure}

\emph{$U(1)$ Gauge Theory}.    We begin by studying a prototypical one-dimensional model exhibiting confinement, namely a $U(1)$ lattice gauge theory.  This model is simply the one-dimensional version of quantum electrodynamics, which can be defined as follows:
\begin{multline}
H_{\textrm{bare}}= g \sum_{J}E_{J}^{2} + U_{1}\sum_{j} (n_{b,j}^{2} + n_{d,j}^{2}) + U_{2} \sum_{j}n_{b,j}n_{d,j}\\ + t\sum_{j}\big[ b_{j}b_{j+1}^{\dagger}e^{-iA_{J}} + d_{j}d_{j+1}^{\dagger}e^{iA_{J}} + \textrm{h.c.} \big],
\label{fullHamiltonian}
\end{multline} 
where $\{j,\ldots \}$ denote sites, and $\{ J, \ldots \}$ represent links (Fig. \ref{fig:schwinger}). The gauge fields $A_{J}$ live on the links, and $E_{J}$ are the canonically conjugate electric fluxes. We have introduced two particle types in this model: positively and negatively charged particles, which are created at site $j$ by the $b_{j}^{\dagger}$ and $d_{j}^{\dagger}$ operators respectively, while $n_{b,j}$ and $n_{d,j}$ denote the corresponding number operators.  These charges obey the Gauss's law constraint $\partial_x E = n_b - n_d$ on every site.  The terms of the first line represent the energy stored in the electric strings (flux lines), as well as intra-species and inter-species on-site repulsions between charges.  The second line represents the coupling between motion of charge and growing/shrinking an electric string.  When the charges are fermions, this model can be viewed as a lattice version of the massive Schwinger model.\cite{schwinger,coleman}

While this Hamiltonian cannot be exactly solved for an arbitrary number of $b$ and $d$ particles, it can be analyzed by perturbatively expanding about $t=0$, at which point there is no dynamics in the system.  We will assume throughout that $t\ll g,U_1,U_2$.  The interaction term for our perturbation theory is then simply the hopping term, and is given by:
\begin{equation}
H_{\textrm{int}}=V= t\sum_{j}\big[ b_{j}b_{j+1}^{\dagger}e^{-iA_{J}} + d_{j}d_{j+1}^{\dagger}e^{iA_{J}} + \textrm{h.c.} \big]
\label{inthamiltonian}
\end{equation}
We will also work in the limit where the $b$ and $d$ charges obey a mutual hard--core repulsion, i.e. we take $U_{2} \rightarrow \infty$.  As we will see, this mutual repulsion plays a crucial role in giving rise to a conserved dipole moment.

\begin{figure}[t]
 \centering
 \includegraphics[width=0.4\textwidth]{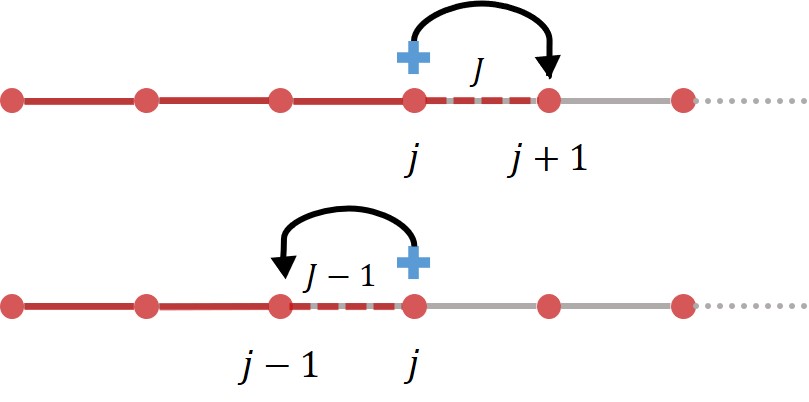}
\captionsetup{justification=raggedright,singlelinecheck=false}
 \caption{Allowed hopping processes in the single-particle sector at second order in perturbation theory.}
 \label{fig:single}
 \end{figure}

To understand the dynamics of this model, we first study the behavior of the single-particle sector.  Consider the initial state to be a $+$ charge at site $j$ with a semi-infinite electric string extending to the left, $i.e.$ $|+_j\rangle = (\prod_{I<j}e^{-iA_I}) b_{j}^{\dagger}|0\rangle$, where $|0\rangle$ is the state with no particles or electric field (vacuum).  We now perform a perturbative analysis within this sector, equivalent to a Schrieffer-Wolff transformation\cite{schriefferwolff,takahashi1977half} (see also Refs. \onlinecite{sous2019fractons} and \onlinecite{slagle1}), integrating out the gauge fields to obtain an effective Hamiltonian for the charges.  We evaluate the effective single-particle Hamiltonian at second order as follows:
\begin{equation}
h^{(2)}_{1} = PV \dfrac{1-P}{\varepsilon_{0}-H_{0}}VP,
\label{pert}
\end{equation}
where $H_{0}$ is the unperturbed Hamiltonian, $\varepsilon_{0}$ is the unperturbed energy, and $P$ is the projector onto states with flux strings of the same length and sign as in the initial state. For the initial state considered, $\varepsilon_{0}=U_{1}$. It is clear that $P$ acting on the initial state returns the same state. Under the action of $V$, there are two possible processes (see Fig. \ref{fig:single}), i.e.
\begin{equation}
V|+_j\rangle = t \big( b_{j}b_{j+1}^{\dagger}e^{-iA_{J}} + b_{j-1}^{\dagger}b_{j}e^{iA_{J-1}}\big)|+_j\rangle
\label{finalState1}
\end{equation} 
$1-P$ acting on this superposition returns the same state. The energy difference from the initial state is $g$ for one intermediate state and $-g$ for the other.  Acting with $PV$ on these states returns $|+_j\rangle$. The effective Hamiltonian, featuring only an onsite energy term and no single-particle hopping is given by:
\begin{equation}
\mathcal{H}_{\textrm{eff}}^{(1)} = U_{1} \sum_{j}n_{b,j}^{2}
\label{eff2nd}
\end{equation}
Note the absence of renormalization of $U_1$, due to the cancellation of signs from the two intermediate states.  At this level of perturbation theory, the charges in the single-particle sector are strictly immobile, consistent with fracton behavior.  (See also Ref. \onlinecite{mazza} for a related discussion on the suppression of single-particle transport.)  It is not hard to see that this immobility holds to all orders in perturbation theory, due to the structure of the projectors.  Importantly, there are no issues associated with the semi-infinite nature of the attached electric string, as we verify below.

  \begin{figure}[b]
 \centering
 \includegraphics[width=0.3\textwidth]{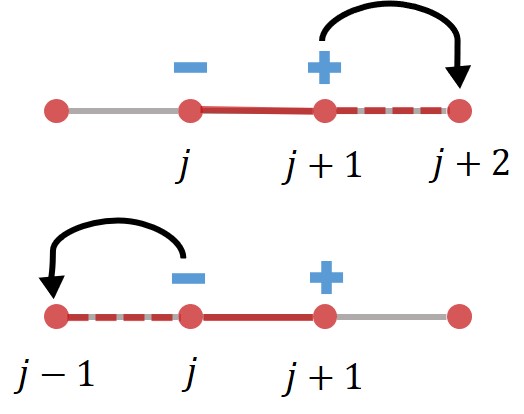}
\captionsetup{justification=raggedright,singlelinecheck=false}
 \caption{Allowed hopping processes in the two-particle sector at second order in perturbation theory, in the limit where the $b$ and $d$ particles obey hard--core repulsion.}
 \label{fig:double}
 \end{figure}
 
To make a more precise connection with fracton physics, we should also examine mobility of dipolar bound states.  To this end, we repeat the above analysis in the two-particle sector, with one positive and one negative charge.  We start with an initial state $e^{-iA_J}b_{j+1}^{\dagger} d_{j}^{\dagger}|0\rangle$.  To second order in perturbation theory, the two-particle Hamiltonian $h_{2}^{(2)}$ is given by the expression in Eqn. \ref{pert}, with $V$ and $P$ defined as in the single-particle case. For this initial state, $\varepsilon_{0} = 2U_{1}$. $P$ returns the same state. Working within the hard--core repulsion ($U_{2} \rightarrow \infty$) limit, we see that there are two possible processes under the action of $V$ (Fig. \ref{fig:double}). $1-P$ trivially acts on these intermediate states. The energy difference between the initial state and the ones shown in Fig. $\ref{fig:double}$ is $-g$. Acting with $PV$ returns the two charges to their original separation, either at their original location or jointly shifted by one site in either direction.  Putting all this together, we obtain the effective Hamiltonian to second order in $t$ to be:
\begin{multline}
\mathcal{H}_{\textrm{eff}}^{(2)} = \bigg(U_{1} - \dfrac{t^{2}}{g} \bigg) \sum_{j} \big( n_{b,j}^{2} + n_{d,j}^{2} \big)\\ - \dfrac{t^{2}}{g} \sum_{j} \big( b_{j+2}^{\dagger}d_{j+1}^{\dagger} + b_{j}^{\dagger}d_{j-1}^{\dagger} \big)b_{j+1}d_{j} 
\label{eff2nd_2particle}
\end{multline}
There are no single-particle hopping processes in this effective Hamiltonian: there is an onsite energy term, and a pair--hopping term which moves a dipole at sites $(j,j+1)$ to either $(j+1,j+2)$ or $(j-1,j)$. Therefore, to second order in perturbation theory, single charges remain immobile, while dipoles move freely, in a manifestation of fracton physics.  Indeed, this effective Hamiltonian explicitly exhibits conservation of the dipole moment:
\begin{equation}
P = \sum_i (n_{b,i} - n_{d,i})\,x_i = \textrm{constant}
\end{equation}
as expected for a fracton system.  Higher order terms in the perturbation series will lead to beyond-nearest-neighbor hopping of the dipoles, but will not violate the conservation of dipole moment.

Besides nearest-neighbor dipoles, it is also important to study the behavior of dipoles with larger separation between the charges, to verify that there is indeed a sensible single-particle limit when the charges are well separated.  Whereas minimal dipoles moved only via intermediate processes which grew the electric string, longer dipoles can move via intermediate steps with a shortened string, leading to fortuitous sign cancellations in the perturbation series.  For example, consider a dipole consisting of a positive and negative charge separated by two sites.  It can readily be checked from Eqn. \ref{pert} that, to second order in perturbation theory, there is exact cancellation between growing and shrinking processes, and no dipole hopping is induced.  Rather, this dipole only moves at fourth order in perturbation theory, $i.e.$ with a hopping matrix element of order $t^4/g^3$.  Similarly, a dipole with charges separated by $n$ sites has hopping matrix elements of order $t^{2n}/g^{2n-1}$.  As long as we remain in the regime $t\ll g$, there will be negligible hopping of well-separated charges, allowing for a sensible single-particle limit of immobile fractons.  This slow-down of separated charges is necessitated by the locality of the model, which leads to retardation of the gauge-mediated interaction between charges.  Such a slow-down would not be manifest if the gauge field were replaced by an instantaneous linear potential between the charges.  We note, however, that an instantaneous linear potential can lead to similar dipole-conserving behavior.\cite{bloch}

\emph{Ising Model}.    Our analysis can also be straightforwardly applied to other confined models in one dimension, such as the quantum Ising chain with transverse and longitudinal fields, as considered in Refs. \onlinecite{crusoe},\onlinecite{crusoe2}:
\begin{equation}
H_{\textrm{Ising}}=\sum_{J} \big( \mathcal{J}\sigma^{z}_{J}\sigma^{z}_{J+1}+h_{x}\sigma^{x}_{J}+h_{z}\sigma^{z}_{J} \big)
\label{Ising}
\end{equation}
where $\sigma^{a}_{J}$ ($a=x,y,z$) are the Pauli matrices acting on the $J^{th}$ link of the chain, $\mathcal{J}$ is the exchange parameter, and $h_{x}$ and $h_{z}$ are the transverse and longitudinal fields respectively.  To connect with the analysis of the previous section, it is simplest to rewrite this model in the language of a $Z_2$ gauge theory.  We therefore introduce redundant ``matter" fields $\tau^{a}_{j}$ ($a=x,y,z$), representing domain walls living on the sites of the chain ($j$ is the site between links $J-1$ and $J$), imposing a Gauss's law constraint given by $\sigma^{z}_{J-1}\tau^{x}_{j}\sigma^{z}_{J}=1$.  The Hamiltonian in Eqn. \ref{Ising} can then be written as:
\begin{equation}
H_{\textrm{Ising}}^{\textrm{matter}}=\sum_{j} \big( \mathcal{J}\tau^{x}_{j}+h_{x}\tau^{z}_{j}\sigma^{x}_{J}\tau^{z}_{j+1}+h_{z}\sigma^{z}_{J} )
\label{IsingMatter}
\end{equation}
As in the $U(1)$ case, we have terms representing the energy cost of electric strings ($h_z$) and of particles ($\mathcal{J}$), along with a term coupling the motion of charges to the growing/shrinking of electric strings ($h_x$).  In other words, the motion of domain walls is tied to the growing/shrinking of domains, as expected.

This Hamiltonian can be perturbatively analyzed around $h_{x}=0$, where there is no dynamics in the system. As before, the interaction term is the hopping term, and is given by
\begin{equation}
H_{\textrm{int}}=V=\sum_{j} h_{x}\tau^{z}_{j}\sigma^{x}_{J}\tau^{z}_{j+1}
\label{int}
\end{equation}
Starting with a $\mathbb{Z}_{2}$ charge at site $j$, we perturbatively analyze this single particle sector by effectively integrating out the gauge fields ($\sigma$'s). The effective single particle Hamiltonian at second order is again given by Eqn. \ref{pert}, with all the variables retaining their meaning from before.  Even in this case we see that the charges in the single-particle sector are strictly immobile, \emph{i.e.} 
\begin{equation}
\mathcal{H}_{\mathbb{Z}_{2},\textrm{eff}}^{(1)} = \mathcal{J} \sum_{j}\tau^{x}_{j}
\label{z2single}
\end{equation}
where once again $\mathcal{J}$ is not renormalized due to a cancellation of signs from intermediate states.
 
Performing a similar analysis in the two--particle sector, starting from a nearest-neighbor pair of charges, we see that the effective Hamiltonian is given by:
\begin{multline}
\mathcal{H}_{\mathbb{Z}_{2},\textrm{eff}}^{(2)} = \bigg( \mathcal{J} - \dfrac{h_{x}^{2}}{h_{z}}+\dfrac{h_{x}^{2}}{2\mathcal{J}+h_z} \bigg) \sum_{j}\tau^{x}_{j} \\ - \dfrac{h_{x}^{2}}{h_{z}}\sum_{j}\tau^{z}_{j-1}\tau^{z}_{j+1}\big(1-\tau^{x}_{j}\big)
\label{z2two}
\end{multline}
The first term is simply the renormalization of the energy per charge.  The second term represents a pair-hopping term, moving a pair of charges at sites $j-1$ and $j$ to sites $j$ and $j+1$.  This can equivalently be thought of as an ``assisted" hopping term, in which a particle at site $j-1$ can hop two sites to $j+1$, but only if site $j$ is occupied by a second particle.  Either way, we see that particles are strictly immobile in isolation, but gain mobility in the vicinity of other particles, in a manifestation of fracton physics.  We note that the formation of mobile bound states of Ising domain walls has also been discussed in the context of a weakly-broken $E_8$ symmetry\cite{kjall}, which has been detected experimentally.\cite{coldea}

\emph{Interpretation via String Motion}.    We have now shown that one-dimensional confining models can naturally give rise to fracton dynamics, where single particles do not move, while two-particle bound states move with a fixed separation.  In any model featuring a binding potential, a two-particle bound state tends to move with relatively fixed separation.  What distinguishes a confining system is that the potential continues to lock the motion of particles together when they are well separated.  Due to the locality of the model, the motion of a pair of particles proceeds more and more slowly as the particles are separated, asymptotically resulting in immobile fracton behavior at large separations.

An intuitive way to understand this behavior is to recall that the charges can simply be regarded as the endpoints of electric strings, which have both mass and internal tension.  The fixed separation between charges in a dipole is due to the large string tension, which seeks to keep the string at a fixed length.  The dipole hopping terms in the effective Hamiltonian have a simple interpretation as motion of segments of strings.  The slow motion of larger dipoles can be naturally understood in terms of the larger mass associated with longer strings.  An isolated fracton corresponds to a semi-infinite string, which has an infinite mass and therefore cannot be moved by any finite force.

This perspective hints at the possibility of higher-dimensional generalizations of our analysis.  In any dimension, confining models are typically characterized by a massive string with large tension binding pairs of particles together.  Based on this, it is tempting to conclude that an isolated particle ($i.e.$ string endpoint) in higher dimensions should also be immobile, due to the mass of its attached semi-infinite string.  (This relies on having an isolated system, without access to a heat bath, such that the string cannot easily relax.)  However, we must also account for the possibility in higher dimensions that a charge could move via a string bending while maintaining a fixed length.  To avoid this type of motion, we propose adding additional terms to the Hamiltonian, such as $U'(\nabla\times E)^2$, which energetically penalize bending of electric strings.  In the limit of large $U'$, strings will be unable to bend, and single charges will be immobile.  Meanwhile, a pair of particles will be able to move maintaining a fixed separation, with the effective mass increasing with the separation.  Note, however, that while the separation between two particles will remain constant, their relative orientation will not, since nothing prevents the string from rotating in higher dimensions.  As such, a higher-dimensional confining model may not possess component-wise conservation of dipole moment, but rather only its magnitude.  A precise formulation of this pseudo-fracton behavior in higher-dimensional confining models is an interesting open question.

\emph{Conclusions}.    In this work, we have demonstrated a close relationship between two previously studied mechanisms for ergodicity breaking in one dimension, namely confinement and fracton behavior. This in turn draws a neat connection between the scarring phenomenon discussed in the confining models of Refs. \onlinecite{crusoe},\onlinecite{crusoe2} and the dynamical scars\cite{fractonscars} observed in fracton systems. By studying one-dimensional lattice gauge theories, we have shown how perturbatively integrating out the confining field can lead to an effective fracton Hamiltonian for the charges, featuring immobile single particles which can form mobile dipoles.  We have provided intuition for these results in terms of the motion of strings, which hints at potential higher-dimensional generalizations.  Our work opens the door for the fruitful exchange of ideas between the fields of fractons and confinement, for example by providing a complementary understanding of nonergodic behavior in confining systems in terms of corresponding results in the fracton literature.  In turn, the connection between confining models and Rydberg atom arrays\cite{lgt,rydberg} makes these results have direct experimental relevance.

\begin{acknowledgments}
We thank John Sous for a prior collaboration which developed some of the techniques used in this work.  We also thank Kevin Slagle for useful feedback on the manuscript.  The work of MP is supported by the Air Force Office of Scientific Research under award number FA9550-17-1-0183.  The work of SP is supported by the U.S. Department of Energy, Office of Science, Basic Energy Sciences (BES) under Award number DE-SC0014415.
\end{acknowledgments}

\bibliography{library}

\end{document}